\documentstyle[preprint,aps]{revtex}

\newcommand{\bb}{\begin{equation}}  \newcommand{\ee}{\end{equation}}
\newcommand{\Sch}{Schwarzschild}    \newcommand{\w}{\mbox{\tiny $\wedge$}}
\newcommand{\br}{\begin{eqnarray}}  \newcommand{\er}{\end{eqnarray}}
\newcommand{\C}{\mbox{$l^{-2}$}}    \newcommand{\m}{\mbox{$\frac{1}{2}$}}

\begin{document}
\draft
\preprint{gr-qc/9309011}

\title{BLACK HOLES IN EINSTEIN-LOVELOCK GRAVITY\footnote{Talk given at
the VIII Latin American Symposium on Relativity and Gravitation,
SILARG, Sao Paulo, July 1993.}}

\author{M\'aximo Ba\~nados}
\address{Centro de Estudios Cient\'{\i}ficos de Santiago, Casilla 16443,
Santiago 9, Chile.}

\maketitle

\begin{abstract}
Static, spherically symmetric solutions of the field equations
for a particular dimensional continuation of general relativity
with negative cosmological constant are found.  In even
dimensions the solution has many similarities with the \Sch\
metric. In odd dimensions, the equations of motion are
explicitly anti de-Sitter invariant, and the solution is alike in
many ways to the 2+1 black hole.
\end{abstract}

This talk will be devoted to the study of lower and higher
dimensional black holes in the Einstein-Lovelock theory of
gravity . The results presented here have been developed in
collaboration with C. Teitelboim and J. Zanelli\cite{BTZ,BTZ2}.
I thank them for their great encouragment and guidance while
this work was in preparation. Of course, errors and omitions in
this report are my responsability.

In dimensions greater than four the usual Einstein equations are
not the most general equations that give rise to second order
tensorial equations for the metric.  In contrast, a large class
of equations, parametrized by a number of independent
dimensionful parameters, can be considered\cite{Lovelock}. This
equations are known as Einstein-Lovelock equations.  In this
work black-hole solutions for a particular class of these
equations will be found.

The construction of the most general set of equations for
gravity in an arbitrary spacetime dimension $D$ is more easily
done in the first order formalism.  In this framework, the
canonical variables are the matrix $e^a_{\mu}$ and the spin
connection $w^a_{\;b \mu}$. These two fields are related with
the metric and the Christoffel symbols by the change of
coordinates

\br
g_{\mu \nu} &=& e^a_{\mu} \, \eta_{ab} \, e^b_{\nu}  \\
\Gamma^{\rho}_{\mu\nu} &=& e^{\rho}_a \, w^a_{\;b\nu} \,
e^b_{\mu} + e^{\rho}_a \, e^a_{\mu,\nu}
\label{6}
\er
where $e^a_{\mu}$ acts as the matrix of change of basis between
the coordinate and orthonormal basis.

Consider the pair $\{ e^a = e^a_{\mu} dx^{\mu}, w^{ab} =
w^{ab}_{\mu}dx^{\mu} \}$ as a connection 1-form for a gauge
theory for the anti-de Sitter group.  This group has generators
$J_a$ and $J_{ab}$ satisfying the commutation relations

\begin{eqnarray}
{}\left[ J_{a} ,J_{b} \right] & = & \C J_{ab} \nonumber \\
{}\left[ J_{ab} ,J_{c} \right] & = & J_{a} \eta_{bc} - J_{b}
  \eta_{ac} \label{7} \\
{}\left[ J_{ab} ,J_{cd} \right] & = & -J_{ac} \eta_{bd} + J_{ad}
  \eta_{bc} +  J_{bc}\eta_{ad} - J_{bd}\eta_{ac} \nonumber
\end{eqnarray}
where $l$ is a parameter with dimensions of a lenght.

Define the 1-form gauge field $A$ as

\bb
A \equiv e^a J_a + \m w^{ab} J_{ab}
\label{8}
\ee
and the 2-form Yang-Mills curvature as

\br
F &=& dA + A \w A \nonumber \\
  &\equiv & F^a J_a + \m F^{ab}J_{ab}
\label{9}
\er
Using the commutation relations (\ref{7}) it is simple to find
$F^a$ and $F^{ab}$ in terms of $e^a$ and $w^{ab}$,

\br
F^a    &=& de^a + w^a_{\;b} \w e^b \label{10} \\
F^{ab} &=& R^{ab} + \C e^a \w e^b \label{11}
\er
where $R^{ab} = dw^{ab} + w^a_c \w w^{cb}$ is the spacetime
curvature tensor.

We will use the anti-de Sitter 2-form curvature $F$ to
write the equations of motion.  Consider first the three
dimensional case. As the lagrangian is a 3-form, and the gauge
field $A$ is a 1-form, the equations of motion must be 2-forms.
By inspection it is easy to check that, if no other fields
are included, the only anti-de Sitter invariant equations that
one can write are

\bb
F=0.
\label{12}
\ee
If Eq. (\ref{12}) is written in terms of the metric using
(\ref{6}), one can easily prove that it is equivalent to the
usual Einstein equation in three dimensions.  Black-hole
solutions for (\ref{12}) exists\cite{BTZ}. The metric takes
the form

\bb
ds^2 = -N^2 dt^2 + N^{-2} dr^2  + r^2(N^{\phi} dt + d\phi)^2
\label{12a}
\ee
where the squared lapse $N^2(r)$ and the angular shift
$N^{\phi}(r)$ are given by

\begin{eqnarray}
N^2(r) &=& - M + \frac{r^2}{l^2} + \frac{J^2}{4r^2} \label{12b} \\
N^{\phi}(r) &=& - \frac{J}{2r^2}  \label{12c}
\end{eqnarray}
with $-\infty < t <\infty$, $0 < r < \infty$ and $0 \leq \phi
\leq 2\pi$.  The parameters $M$ and $J$ are, respectively, the
mass and the angular momentum of the solution.

An interesting property of three dimensions is that a
general solution for the equations (\ref{12}) can be given. In
fact, the expression

\bb
A = U^{-1} dU.
\label{13}
\ee
where $U$ is an element of the gauge group is the most general
solution for (\ref{12}).  The group element $U$ that reproduces
the black-hole metric (\ref{12a}) was found by Cangemi et
al.\cite{Cangemi}.  Of course, $U$ is not trivial in the sense
that it cannot be continuosly deformed to the identidy. (If this
can be done, $A$ would not represent a physically interesting
solution since it can be set equal to zero by a gauge
transformation.)

{}From the point of view of the metric, the non-trivial topology
of the manifold is seen from the fact that the metric
(\ref{12a}) can be obtained from anti-de Sitter space by means
of an identification with a discrete subgroup of the symmetry
group. This identification changes the topology of anti de-Sitter
space and non-contractible loops appear\cite{BHTZ}.

Let us now study the higher odd-dimensional cases. In all odd
dimensions the equations of motion must be a $(2n-2)$-form where
$2n-1$ is the corresponding spacetime dimension.  As the only
tensor in the theory (without including external fields), is the
curvature 2-form $F$ it is not difficult to check that the only
possible equations are

\bb \begin{array}{rcl}
           F \w F &=& 0 \mbox{~~~~~~~~}D=5 \\
      F \w F \w F &=& 0 \mbox{~~~~~~~~}D=7 \\
                  &\vdots &      \\
 F \w F \w...\w F &=& 0 \mbox{~~~~~~~~}D=2n-1
\end{array}  \label{14}
\ee
It is remarkable that, although the equations (\ref{14}) are
much more complicated than the case $F=0$, it is still
possible to solve them in closed form for a spherically
symmetric static metric.  For any odd dimension $D=2n-1$
the black hole metric

\bb
ds^2 = - \left[1 - (M+1)^{\frac{1}{n-1}} + (r/l)^2\right]dt^2 +
\frac{dr^2}{1 - (M+1)^{\frac{1}{n-1}} + (r/l)^2} + r^2 d\Omega^2
\label{15}
\ee
is an exact solution of Eqs. (\ref{14}).

This geometry represents a natural extension of the three
dimensional case (without angular momentum). A horizon exists
only for positive masses and, in the special case $M=-1$, the
solution reduces to anti-de Sitter space.  An important
difference with the three dimensional case is that the curvature
tensor for $D>3$ is no longer constant. Moreover,
the curvature is singular at the origin although its effects are
not observed from outside due to the presence of the horizon.

In the even dimensional case it is not possible to write a set
of equations invariant under de anti-de Sitter group. This is
simply observed from the fact that in even dimensions, the
equations must be forms of odd degree and there are no odd
tensor forms in the theory. It is necessary to break the
symmetry under the full group. The best alternative is to
consider a set of equations invariant only under the Lorentz
group as it is a subgroup of the anti-de Sitter group.  (That
can be seen from the fact that the $J_{ab}$ form a subalgebra in
(\ref{7}).) Thus, besides the curvature 2-form $F$, one has
the 1-form $e^a$ that transforms as a vector under
Lorentz transformations.  In four dimensions the simplest
equation that one can write is

\bb
F \w e = 0
\label{16}
\ee
and it is a simple matter to check that (\ref{16}) are
equivalent to the
usual Einstein equations with cosmological constant. In
higher dimensions, a natural choice for the remaining
equations is, in view of (\ref{14}),

\bb \begin{array}{rcl}
           F \w F \w e &=& 0 \mbox{~~~~~~~~}D=6 \\
      F \w F \w F \w e &=& 0 \mbox{~~~~~~~~}D=8 \\
                       &\vdots &      \\
F \w F \w ...\w F \w e &=& 0 \mbox{~~~~~~~~}D=2n
\end{array}  \label{17}
\ee
[Note that the equations (\ref{17}) are the dimensionally
continued version of the equations (\ref{14}).]

Again, these equations can be solved in the spherically
symmetric case obtaining the black hole geometry

\bb
ds^2 = -\left[1 - (2M/r)^{\frac{1}{n-1}} + (r/l)^2\right]dt^2 +
\frac{dr^2}{1 - (2M/r)^{\frac{1}{n-1}} + (r/l)^2} + r^2
d\Omega^2.
\label{18}
\ee
This metric represents a natural extension for the \Sch\
solution. There exists a horizon only for positive masses.
Anti-de Sitter space, in this case, has zero mass. The curvature
tensor is singular at the origin for non-zero mass but
this singularity is hidden by the horizon.

It should be stressed here that the equations of motion written
in (\ref{17}) are not the most general equations that can be
considered without breaking the Lorentz symmetry. For example,
in six dimensions one can consider

\bb
F \w F \w e + a F \w e \w e \w e + b e \w e \w e \w e \w e =0
\label{19}
\ee
where $a$ and $b$ are arbitrary parameters. In this more
general case the solution can still be
found, but it cannot be written explicitly as the metric
components are the roots of algebraic equations that cannot be
solved in a unique way. Our case, which corresponds to take

\bb
a = b = 0,
\label{20}
\ee
can be thought of as the dimensionally continued version of
equations (\ref{14}) for any even dimension. It is interesting
to note that in the odd dimensional case, if one break down the
symmetry to the Lorentz group, then the same arbitrariness in
the coefficients appear. However, in odd dimensions
there exist the priviliged case with a larger symmetry that fixes
the coefficients.

Note that the equations (\ref{14}) and (\ref{17}) do not
coincide with the usual dimensionally continued Einstein
equations which are linear in the curvature tensor,

\bb
F \w \underbrace{e \w \cdots \w e}_{D-3} = 0.
\label{21}
\ee
It is only in the special cases of $D=3,4$ when (\ref{21})
coincides with (\ref{12}) and (\ref{16}).

The black hole solutions (\ref{15}) and (\ref{18}) admit
electric charge. We will not write this solution here but only
mention that a new curvature singularity develops, although it
is also hidden by the horizon\cite{BTZ2}.  These black holes
have thermodynamical properties similar to those in three and
four dimensions. The temperature goes to infinity as the black
hole dissapears in all even dimensions while in the odd
dimensional case the temperature goes to zero. The entropy, on
the other hand, is no longer proportional to the area of the
black hole but it is still an increasing function of the black
hole radius and mass\cite{BTZ2}.\\

This work was partially supported by the grant 193.0910/93 from
FONDECYT (Chile), by a European Communities research contract,
and by institutional support to the Centro de Estudios
Cient\'{\i}ficos de Santiago, provided by SAREC (Sweden) and a
group of chilean private companies (COPEC, CMPC, ENERSIS).


\begin{thebibliography}{10}
\bibitem{BTZ}  M. Ba\~nados, C. Teitelboim and J. Zanelli, {\em
  Phys. Rev. Lett.}, {\bf 69}(1992)1849.
\bibitem{BTZ2} M. Ba\~nados, C. Teitelboim and J. Zanelli,
CECS/IAS preprint (1993), gr-qc/9307033, to appear in {\em Phys.
               Rev. D.}
\bibitem{Lovelock} D. Lovelock, {\em J. of Math. Phys.} {\bf
                  12}(1971)498.
\bibitem{BHTZ} M. Ba\~nados, M. Henneaux, C. Teitelboim and J.
Zanelli, CECS/IAS-preprint (1992), gr-qc/9302012, to appear in
            {\em Phys. Rev. D.}
\bibitem{Cangemi} D. Cangemi, M. Leblanc and R.B. Mann,
            CTP\#2162 preprint, to appear in {\em Phys. Rev. D.}
\end{thebibliography}
\end{document}